\documentclass{llncs}

\usepackage[labelfont=bf]{caption}

\usepackage[utf8]{inputenc}
\usepackage{cite}
\usepackage{subcaption}
\usepackage{wrapfig}

\usepackage{xcolor}

\usepackage{dsfont}

\usepackage{xspace}
\usepackage{enumerate}

\usepackage{microtype}

\usepackage{amsmath}
\usepackage{amssymb}
\usepackage{amstext}
\usepackage{esvect}
\usepackage{amsopn}

\usepackage{hyperref}
\usepackage{graphicx}
\usepackage[noline,boxruled, noend,algonl,boxed]{algorithm2e}
\usepackage[basic]{complexity}

\usepackage{paralist}
\defaultleftmargin{0.0em}{0.0em}{0.0em}{0.0em}
\usepackage{enumitem}

\usepackage[left, displaymath, mathlines]{lineno}
\newcommand*\patchAmsMathEnvironmentForLineno[1]{%
  \expandafter\let\csname old#1\expandafter\endcsname\csname #1\endcsname
  \expandafter\let\csname oldend#1\expandafter\endcsname\csname end#1\endcsname
  \renewenvironment{#1}%
     {\linenomath\csname old#1\endcsname}%
     {\csname oldend#1\endcsname\endlinenomath}}%
\newcommand*\patchBothAmsMathEnvironmentsForLineno[1]{%
  \patchAmsMathEnvironmentForLineno{#1}%
  \patchAmsMathEnvironmentForLineno{#1*}}%
\AtBeginDocument{%
\patchBothAmsMathEnvironmentsForLineno{equation}%
\patchBothAmsMathEnvironmentsForLineno{align}%
\patchBothAmsMathEnvironmentsForLineno{flalign}%
\patchBothAmsMathEnvironmentsForLineno{alignat}%
\patchBothAmsMathEnvironmentsForLineno{gather}%
\patchBothAmsMathEnvironmentsForLineno{multline}%
}

\usepackage[textsize=footnotesize]{todonotes}
\SetVlineSkip{0pt}

\def\comment#1{}%
\def\withcomments{%
  \newcounter{mycommentcounter}%
   \def\comment##1{\refstepcounter{mycommentcounter}%
    \ifhmode%
     \unskip%
     {\dimen1=\baselineskip \divide\dimen1 by 2 %
       \raise\dimen1\llap{\tiny
	{-\themycommentcounter-}}}\fi%
     \marginpar[{\renewcommand{\bvarphiaselinestretch}{0.8}%
       \hspace*{-2em}\begin{minipage}{1.5\marginparwidth}\footnotesize%
[\themycommentcounter]:%
\raggedright ##1\end{minipage}}]{\renewcommand{\baselinestretch}{0.8}%
       \begin{minipage}{1.5\marginparwidth}\footnotesize%
[\themycommentcounter]: \raggedright%
##1\end{minipage}}}%
  }

\renewcommand{\figurename}{Fig.}

\renewcommand{\C}{\ensuremath{\mathcal C}}
\newcommand{\U}{\ensuremath{\mathcal U}}

\newcommand{\rephrase}[3]{\noindent\textbf{#1~#2.}~\emph{#3}}

\newcounter{condition} 


\makeatletter
\let\@mkboth\@gobbletwo
\def\@oddhead{\small\rm
\hfil
{Ordered Level Planarity, Geodesic Planarity and Bi-Monotonicity
}
\rightmark
\qquad\thepage}%
\def\@oddfoot{}\def\@evenhead{\small \rm\thepage\qquad
\leftmark
{~~Boris Klemz and G\"unter Rote
}\hfil}%
\def\@evenfoot{}
\makeatother

\spnewtheorem*{prprty}{Property}{\bfseries}{\itshape}

\spnewtheorem{observation}{Observation}{\bfseries}{\itshape}
\newenvironment{proofsketch}{\noindent{\it Proof Sketch.}}{\hfill $\qed$}

\title{Ordered Level Planarity, Geodesic Planarity and Bi-Monotonicity
\thanks{Due to space constraints, some proofs in the first 12 pages of this manuscript are only sketched or omitted entirely.  Full proofs of all claims can be found in the appendix.}
}
 
\author{Boris Klemz, Günter Rote}

\authorrunning{Klemz and Rote}

\institute{Institute of Computer Science, Freie Universit\"at Berlin, Germany
}

\begin{document}

\ifnum\mag=1200
\voffset=-1,8true cm
\hoffset=-2,5true cm
\pdfpageheight=29.7 true cm
\pdfpagewidth=21 true cm
\fi

\maketitle

\begin{abstract}
We introduce and study the problem \textsc{Ordered Level Planarity} which asks for a planar
drawing of a graph such that vertices are placed at prescribed positions in the plane and such that every edge is realized as a $y$-monotone curve. This can be interpreted as a variant of \textsc{Level Planarity} in which the
vertices on each level appear in a prescribed total order.
We establish a complexity dichotomy with respect to both the maximum degree and the level-width, that is, the maximum number of vertices that share a level.
Our study of \textsc{Ordered Level Planarity} is motivated by connections to several other graph drawing problems.

\textsc{Geodesic Planarity} asks for a planar drawing of a graph such that  vertices are placed at prescribed positions in the plane and such that every edge~$e$ is realized as a polygonal path~$p$ composed of line segments with two adjacent directions from a given set $S$ of directions symmetric with respect to the origin. Our results on \textsc{Ordered Level Planarity} imply $\mathcal{NP}$-hardness for any~$S$ with~$|S|\ge 4$ even if the given graph is a matching. 
Katz, Krug, Rutter and Wolff claimed that for matchings \textsc{Manhattan Geodesic Planarity}, the case where~$S$ contains precisely the horizontal and vertical directions, can be solved in polynomial time \lbrack GD'09\rbrack.
Our results imply that this is incorrect unless~$\mathcal P=\mathcal NP$.
Our reduction extends to settle the complexity of the \textsc{Bi-Monotonicity} problem, which was proposed by Fulek, Pelsmajer, Schaefer and \v Stefankovi\v c.

\textsc{Ordered Level Planarity} turns out to be a special case of \textsc{T-Level Planarity}, \textsc{Clustered Level Planarity} and \textsc{Constrained Level Planarity}. Thus, our results strengthen previous hardness results. In particular, our reduction to \textsc{Clustered Level Planarity} generates instances with only two non-trivial clusters. This answers a question posed by Angelini, Da Lozzo, Di Battista, Frati and Roselli.
\end{abstract}

\section{Introduction}

In this paper we introduce \textsc{Ordered Level Planarity} and study  its complexity. We establish connections to several other graph drawing problems, which we survey in this first section. We proceed from general problems to more and more constrained ones.

\textbf{Upward Planarity:} An \emph{upward} planar drawing of a directed graph is a plane drawing where every edge~$e=(u,v)$ is realized as a~$y$-monotone curve that goes upward from~$u$ to~$v$. Such drawings provide a natural way of visualizing a partial order on a set of items. The problem \textsc{Upward Planarity} of testing whether a directed graph has an upward planar drawing is~$\mathcal{NP}$-complete~\cite{DBLP:journals/siamcomp/GargT01}. However, if the~$y$-coordinate of each vertex is prescribed, the problem can be solved in polynomial time~\cite{DBLP:conf/gd/JungerLM98}. This is captured by the notion of level graphs.

\textbf{Level Planarity:} A \emph{level graph}~$\mathcal G=(G,\gamma)$ is a directed graph~$G=(V,E)$ together with a \emph{level assignment} $\gamma :V\rightarrow \lbrace 0,\dots ,h\rbrace$ where~$\gamma$ is a surjective map with~$\gamma (u)<\gamma (v)$ for every edge~$(u,v)\in E$. Value~$h$ is the \emph{height} of~$\mathcal G$. The vertex set $V_i=\lbrace v\mid \gamma (v)=i\rbrace$ is called the~$i$-th \emph{level} of~$\mathcal G$ and~$\lambda_i=|V_i|$ is its \emph{width}. The \emph{level-width}~$\lambda$ of~$\mathcal G$ is the maximum width of any level in~$\mathcal G$.
A \emph{level} planar drawing of~$\mathcal G$ is an upward planar drawing of~$G$ where the $y$-coordinate of each vertex~$v$ is~$\gamma (v)$. The horizontal line with~$y$-coordinate~$i$ is denoted by~$L_i$. The problem \textsc{Level Planarity} asks whether a given level graph has a level planar drawing.
The study of the complexity of \textsc{Level Planarity} has a long history \cite{DBLP:journals/tsmc/BattistaN88,DBLP:conf/gd/HeathP95, DBLP:conf/gd/JungerLM97,DBLP:conf/gd/JungerLM98,fulek2013hanani}, culminating in a linear-time approach~\cite{DBLP:conf/gd/JungerLM98}.
  \textsc{Level Planarity} has been extended to drawings of level graphs on surfaces different from the plane such as standing cylinder, a rolling cylinder or a torus~\cite{DBLP:conf/gd/AngeliniLBFPR16,DBLP:journals/jgaa/BachmaierBF05,DBLP:conf/esa/BachmaierB08}.

An important special case are  \emph{proper} level graphs, that is, level graphs in which~$\gamma (v)=\gamma (u)+1$ for every edge~$(u,v)\in E$. Instances of \textsc{Level Planarity} can be assumed to be proper without loss of generality by subdividing long edges~\cite{DBLP:conf/gd/JungerLM98,DBLP:journals/tsmc/BattistaN88}. However,
in variations of \textsc{Level Planarity} where we 
 impose additional constraints, the assumption that instances are proper can have a strong impact on the complexity of the respective problems~\cite{DBLP:journals/tcs/AngeliniLBFR15}.

\textbf{Level Planarity with Various Constraints:}
\textsc{Clustered Level Planarity} is a combination of \textsc{Cluster Planarity} and \textsc{Level Planarity}. The task is to find a level planar drawing while simultaneously visualizing a given cluster hierarchy according to the rules of \textsc{Cluster Planarity}. The problem is $\mathcal{NP}$-complete in general~\cite{DBLP:journals/tcs/AngeliniLBFR15}, but efficiently solvable for proper instances~\cite{DBLP:conf/sofsem/ForsterB04,DBLP:journals/tcs/AngeliniLBFR15}.

\textsc{T-Level Planarity} is a consecutivity-constrained version of \textsc{Level Planarity}: every level~$V_i$ is equipped with a tree~$T_i$ whose set of leaves is~$V_i$. For every inner node~$u$ of~$T_i$ the leaves of the subtree rooted at~$u$ have to appear consecutively along~$L_i$. The problem is $\mathcal{NP}$-complete in general~\cite{DBLP:journals/tcs/AngeliniLBFR15}, but efficiently solvable for proper instances~\cite{DBLP:journals/dam/WotzlawSP12,DBLP:journals/tcs/AngeliniLBFR15}. The precise definitions and a longer discussion about the related work are deferred to Appendix~\ref{sec:clusterAndT}.

Very recently, Brückner and Rutter~\cite{DBLP:conf/soda/BrucknerR17} explored a variant of \textsc{Level Planarity} in which the left-to-right order of the vertices on each level has to be a linear extension of a given partial order. They refer to this problem as  \textsc{Constrained Level Planarity} and they provide an efficient algorithm for single-source graphs and show~$\mathcal{NP}$-completeness of the general case.

\textbf{A Common Special Case - Ordered Level Planarity:}
We introduce a natural variant of \textsc{Level Planarity} that specifies a total order for the vertices on each level. An \emph{ordered} level graph~$\mathcal G$ is a triple~$(G=(V,E),\gamma,\chi)$ where~$(G,\gamma)$ is a level graph and $\chi :V\rightarrow \lbrace 0,\dots ,\lambda -1\rbrace$ is a \emph{level ordering} for~$G$. We require that $\chi$ restricted to domain $V_i$ bijectively maps to $\lbrace 0,\dots, \lambda_i-1\rbrace $.
An \emph{ordered} level planar drawing of an ordered level graph~$\mathcal G$ is a level planar drawing of~$(G,\gamma)$ where for every~$v\in V$ the $x$-coordinate of~$v$ is~$\chi (v)$. Thus, the position of every vertex is fixed. The problem \textsc{Ordered Level Planarity} asks whether a given ordered level graph has an ordered level planar drawing.

In the above definitions, the~$x$- and~$y$-coordinates assigned via~$\chi$ and~$\gamma$ merely act as a convenient way to encode total and partial orders respectively. In terms of realizability, the problems are equivalent to generalized versions where~$\chi$ and~$\gamma$ map to the reals. In other words, the fixed vertex positions can be any points in the plane. All reductions and algorithms in this paper carry over to these generalized versions, if we pay the cost for presorting the vertices according to their coordinates. \textsc{Ordered Level Planarity} is also equivalent to a relaxed version where we only require that the vertices of each level~$V_i$ appear along~$L_i$ according to the given total order without insisting on specific coordinates. We make use of this equivalence in many of our figures for the sake of visual clarity.

\textbf{Geodesic Planarity:}
Let~$S\subset \mathbb Q^2$ be a finite set of directions symmetric with respect to the origin, i.e.~for each direction~$s\in S$, the reverse direction~$-s$ is also contained in~$S$. A plane drawing of a graph is \emph{geodesic} with respect to~$S$ if every edge is realized as a polygonal path~$p$ composed of line segments with two adjacent directions from $S$. Two directions of~$S$ are \emph{adjacent} if they appear consecutively in the projection of~$S$ to the unit circle. Such a path~$p$ is  a geodesic with respect to some polygonal norm that corresponds to~$S$.
An instance of the decision problem \textsc{Geodesic~Planarity} is a 4-tuple $\mathcal G=(G=(V,E),x,y,S)$ where~$G$ is a graph, $x$ and $y$ map from~$V$ to the reals and~$S$ is a set of directions as stated above. The task is to decide whether~$\mathcal G$ has a \emph{geodesic drawing}, that is,~$G$ has a geodesic drawing with respect to~$S$ in which every vertex~$v\in V$ is placed at~$(x(v),y(v))$.

Katz, Krug, Rutter and Wolff~\cite{DBLP:conf/gd/KatzKRW09} study \textsc{Manhattan Geodesic Planarity}, which is the special case of \textsc{Geodesic Planarity} where the set~$S$ consists of the two horizontal and the two vertical directions.
Geodesic drawings with respect to this set of direction are also referred to as orthogeodesic drawings~\cite{DBLP:journals/jda/GiacomoGKLR12,DBLP:journals/comgeo/GiacomoFFGK13}.
Katz et al.~\cite{DBLP:conf/gd/KatzKRW09} show that a variant of \textsc{Manhattan Geodesic Planarity} in which the drawings are restricted to the integer grid is~$\mathcal{NP}$-hard even if~$G$ is a perfect matching. The proof is by reduction from \textsc{3-Partition} and makes use of the fact the number of edges that can pass between two vertices on a grid line is bounded. In contrast, they claim that the standard version of \textsc{Manhattan Geodesic Planarity} is polynomial-time solvable for perfect matchings~~\cite[Theorem~5]{DBLP:conf/gd/KatzKRW09}. 
To this end, they sketch a plane sweep algorithm that maintains a linear order among the edges that cross the sweep line. When a new edge is encountered it is inserted as low as possible subject to the constraints implied by the prescribed vertex positions. When we asked the authors for more details, they informed us that they are no longer convinced of the correctness of their approach. 
Theorem~\ref{theorem:lgp} of our paper implies that the approach is indeed incorrect  unless~$\mathcal P=\mathcal{NP}$.

\textbf{Bi-Monotonicity:} Fulek, Pelsmajer, Schaefer and \v Stefankovi\v c~\cite{fulek2013hanani} present a Hanani-Tutte theorem for \emph{y-monotone} drawings, that is, upward drawings in which all vertices have distinct $y$-coordinates. They accompany their result with a simple and efficient  algorithm for \textsc{Y-Monotonicity}, which is equivalent to \textsc{Level Planarity} restricted to instances with level-width~$\lambda=1$. They propose the problem \textsc{Bi-Monotonicity} and leave its complexity as an open problem. The input of \textsc{Bi-Monotonicity} is a triple~$\mathcal G=(G=(V,E),x,y)$ where~$G$ is a graph and~$x$ and~$y$ injectively map from~$V$ to the reals. The task is to decide whether~$\mathcal G$ has a bi-monotone drawing, that is, a plane drawing in which edges are realized as curves that are both $y$-monotone and $x$-monotone and in which every vertex~$v\in V$ is placed at~$(x(v),y(v))$.

\textbf{Main results:}
In Section~\ref{sec:olp} we study the complexity of \textsc{Ordered Level Planarity}. While \textsc{Upward Planarity} is~$\mathcal{NP}$-complete~\cite{DBLP:journals/siamcomp/GargT01} in general but becomes polynomial-time solvable~\cite{DBLP:conf/gd/JungerLM98} for prescribed $y$-coordinates, we show that prescribing both $x$-coordinates and $y$-coordinates renders the problem $\mathcal{NP}$-complete. We complement our result with efficient approaches for some special cases of ordered level graphs and, thereby, establish a complexity dichotomy with respect to the level-width and the maximum degree.

\newcommand{\theoremTwoOlpNpc}{
\textsc{Ordered Level Planarity} is $\mathcal{NP}$-complete, even for maximum degree~$\Delta=2$ and level-width $\lambda=2$. For level-width $\lambda =1$ or~$\Delta^+=\Delta^-=1$ or proper instances \textsc{Ordered Level Planarity} can be solved in linear time, where~$\Delta^+$ and~$\Delta^-$ are the maximum in-degree and out-degree respectively.}
\begin{theorem}
\label{theorem:2olp-npc}
\theoremTwoOlpNpc
\end{theorem}

\textsc{Ordered Level Planarity} restricted to instances with~$\lambda =2$ and~$\Delta =2$ is an elementary problem.
We expect that it may serve as a suitable basis for future reductions.
As a proof of concept, the remainder of this paper is devoted to establishing connections between \textsc{Ordered Level Planarity} and several other graph drawing problems. Theorem~\ref{theorem:2olp-npc} serves as our key tool for settling their complexity. In Section~\ref{sec:lgp} we study \textsc{Geodesic Planarity} and obtain:

\newcommand{\theoremLgp}{\textsc{Geodesic Planarity} is $\mathcal{NP}$-hard for any set of directions $S$ with $|S|\ge 4$ even for perfect matchings in general position.}
\begin{theorem}
\label{theorem:lgp}
\theoremLgp
\end{theorem}

Observe the aforementioned discrepancy between Theorem~\ref{theorem:lgp} and the claim by Katz et al.~\cite{DBLP:conf/gd/KatzKRW09} that \textsc{Manhattan Geodesic Planarity} for perfect matchings is in~$\mathcal P$.
\textsc{Bi-Monotonicity} is closely related to a special case of \textsc{Manhattan Geodesic Planarity}. With a  simple corollary we settle the complexity of \textsc{Bi-Monotonicity} and, thus, answer the open question by Fulek et al.~\cite{fulek2013hanani}.

\newcommand{\theoremBiMon}{\textsc{Bi-Monotonicity} is $\mathcal{NP}$-hard even for perfect matchings.}
\begin{theorem}
\label{theorem:BiMon}
\theoremBiMon
\end{theorem}

\textsc{Ordered Level Planarity} is an immediate and very constrained special case of \textsc{Constrained Planarity}. Further,
in Appendix~\ref{sec:clusterAndT} we establish \textsc{Ordered Level Planarity} as a special case of both \textsc{Clustered Level Planarity} and \textsc{T-Level Planarity} by providing the following reductions.

\newcommand{\theoremTlp}{\textsc{Ordered Level Planarity} with maximum degree~$\Delta =2$ and level-width $\lambda =2$ reduces in linear time to \textsc{T-Level Planarity} with maximum degree $\Delta' =2$ and level-width $\lambda' =4$.}
\begin{theorem}
\label{theorem:olp-to-tlp}
\theoremTlp
\end{theorem}

\newcommand{\theoremClp}{\textsc{Ordered Level Planarity} with maximum degree~$\Delta =2$ and level-width $\lambda =2$ reduces in quadratic time to \textsc{Clustered Level Planarity} with maximum degree $\Delta' =2$, level-width~$\lambda' =2$ and $\kappa'=3$ clusters.}
\begin{theorem}
\label{theorem:olp-to-clp}
\theoremClp
\end{theorem}

Angelini, Da Lozzo, Di Battista, Frati and Roselli~\cite{DBLP:journals/tcs/AngeliniLBFR15} propose the complexity of \textsc{Clustered Level Planarity} for clustered level graphs with a flat cluster hierarchy as an open question.
Theorem~\ref{theorem:olp-to-clp} answers this question by showing that $\mathcal{NP}$-hardness holds for instances with only two non-trivial clusters.

\section{Geodesic Planarity and Bi-Monotonicity}
\label{sec:lgp}

In this section we establish that deciding whether an instance $\mathcal G=(G,x,y,S)$ of \textsc{Geodesic Planarity} has a geodesic drawing is $\mathcal{NP}$-hard even if~$G$ is a perfect matching and even if the coordinates assigned via~$x$ and~$y$ are in \emph{general position}, that is, no two vertices lie on a line with a direction from~$S$. The $\mathcal{NP}$-hardness of \textsc{Bi-Monotonicity} for perfect matchings follows as a simple corollary. Our results are obtained via a reduction from \textsc{Ordered Level Planarity}.

\newcommand{\lemmaOlpToLgp}{
Let~$S\subset \mathbb Q^2$ with $|S|\ge 4$ be a finite set of directions symmetric with respect to the origin. \textsc{Ordered Level Planarity} with maximum degree~$\Delta =2$ and level-width~$\lambda =2$ reduces to \textsc{Geodesic Planarity} such that the resulting instances are in general position and consist of a perfect matching and direction set $S$. The reduction can be carried out using a linear number of arithmetic operations.}
\begin{lemma}
\label{lemma:olp-to-lgp}
\lemmaOlpToLgp
\end{lemma}

\begin{proofsketch}
In this sketch, we prove our claim only for the classical case that $S$ contains exactly the four horizontal and vertical directions.
Our reduction is carried out in two steps.
Let $\mathcal G_o=(G_o=(V,E),\gamma,\chi)$ be an \textsc{Ordered Level Planarity} instance with maximum degree~$\Delta =2$ and level-width~$\lambda =2$. 
In Step (i) we turn $\mathcal G_o$ into an equivalent \textsc{Geodesic Planarity} instance $\mathcal G_g'=(G_o,x',\gamma,S)$.
In Step (ii) we transform $\mathcal G_g'$ into an equivalent \textsc{Geodesic Planarity} instance $\mathcal G_g=(G_g,x,y,S)$ where~$G_g$ is a perfect matching and the vertex positions assigned via $x$ and $y$ are in general position.

\textbf{Step (i):}
In order to transform $\mathcal G_o$ into $\mathcal G_g'$ we apply a shearing transformation. We translate the vertices of each level $V_i$ by $3i$ units to the right, see Figure~\ref{fig:olp-to-gp01}(a) and Figure~\ref{fig:olp-to-gp01}(b). Clearly, every geodesic drawing of~$\mathcal G_g'$ can be turned into an ordered level planar drawing of~$\mathcal G_o$. On the other hand, consider an ordered level planar drawing~$\Gamma_o$ of~$\mathcal G_o$. Without loss of generality we can assume that in~$\Gamma_o$ all edges are realized as polygonal paths in which bend points occur only on the horizontal lines~$L_i$  through the levels~$V_i$ where $0\le i\le h$. Further, we may assume that all bend points have $x$-coordinates in the open interval~$(-1,2)$. We shear~$\Gamma_o$ by translating the bend points and vertices of level $V_i$ by $3i$ units to the right for $0\le i\le h$, see Figure~\ref{fig:olp-to-gp01}(b). In the resulting drawing~$\Gamma_o'$, the vertex positions match those of~$\mathcal G_g'$. Furthermore, all edge-segments have a positive slope.
 Thus, since the maximum degree is~$\Delta =2$ we can replace all edge-segments with $L_1$-geodesic rectilinear paths that closely trace the segments and we obtain a geodesic drawing~$\Gamma_g'$ of~$\mathcal G_g'$, see Figure~\ref{fig:olp-to-gp01}(c).
\begin{figure}[tb]
  \centering
    \includegraphics[width=0.99\columnwidth]{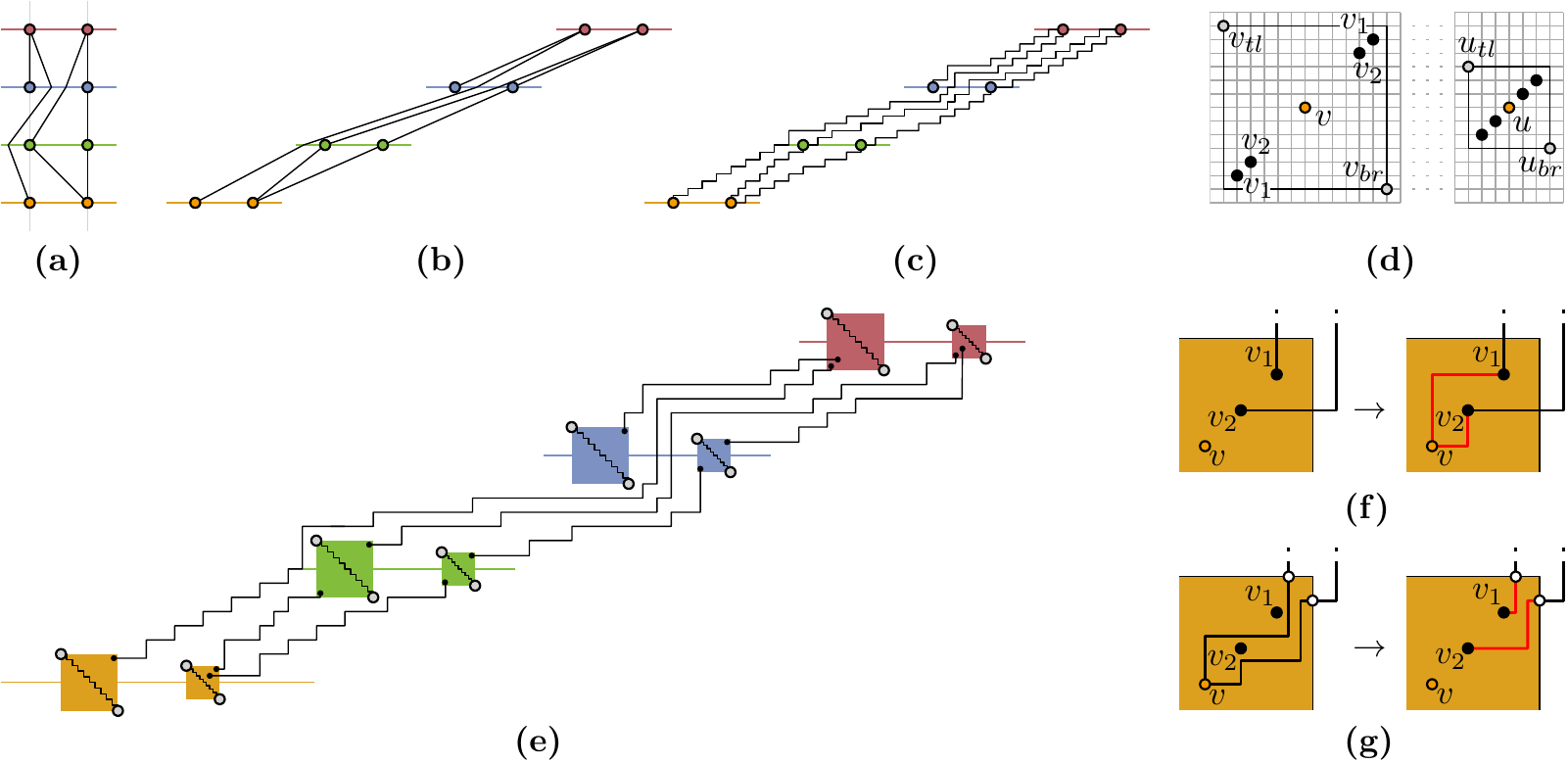}
  \caption{\textbf{(a)}, \textbf{(b)} and \textbf{(c)}: Illustrations of Step (i). \textbf{(d)} The two gadget squares of each level. Grid cells have size $1/48\times 1/48$. \textbf{(e)} Illustration of Step (ii). Turning a drawing of $\mathcal G_g$ into a drawing of $\mathcal G_g'$ \textbf{(f)} and vice versa \textbf{(g)}.}
  \label{fig:olp-to-gp01}
\end{figure}

\textbf{Step (ii):}
In order to turn~$\mathcal G_g'=(G_o=(V,E),x',\gamma,S)$ into the equivalent instance $\mathcal G_g=(G_g,x,y,S)$ we transform $G_o$ into a perfect matching. To this end, we split each vertex $v\in V$ by replacing it with a small gadget that fits inside a square~$r_v$ centered on the position~$p_v=(x'(v),\gamma (v))$ of~$v$, see Figure~\ref{fig:olp-to-gp01}(e). We call~$r_v$ the \emph{square} of~$v$ and use $p_v^{tr}$, $p_v^{tl}$, $p_v^{br}$ and $p_v^{bl}$ to denote the top-right, top-left, bottom-right and bottom-left corner of~$r_v$, respectively.
We use two different sizes to ensure general position.
The size of the gadget square is $1/4\times 1/4$ if $\chi (v)=0$ and it is $1/8\times 1/8$ if~$\chi (v)=1$. The gadget contains a degree-1 vertex for every edge incident to~$v$. In the following we explain the gadget construction in detail, for an illustration see Figure~\ref{fig:olp-to-gp01}(d). Let~$\lbrace v,u\rbrace$ be an edge incident to~$v$. We create an edge~$\lbrace v_1,u\rbrace$ where~$v_1$ is a new vertex which is placed at $p_v^{tr}-(1/48,1/48)$ if~$u$ is located to the top-right of~$v$ and it is placed at~$p_v^{bl}+(1/48,1/48)$ if~$u$ is located to the bottom-left of~$v$. Similarly, if~$v$ is incident to a second edge~$\lbrace v,u'\rbrace$, we create an edge~$\lbrace v_2,u'\rbrace$ where~$v_2$ is placed at $p_v^{tr}-(1/24,1/24)$ or $p_v^{bl}+(1/24,1/24)$ depending on the position of~$u'$. Finally, we create a \emph{blocking} edge~$\lbrace v_{tl},v_{br}\rbrace$ where~$v_{tl}$ is placed at~$p_v^{tl}$ and~$v_{br}$ is placed at~$p_v^{br}$. The thereby assigned coordinates are in general position and the construction can be carried out in linear time.

Assume that~$\mathcal G_g$ has a geodesic drawing~$\Gamma_g$. By construction, all blocking edges have a top-left and a bottom-right endpoint. On the other hand, all other edges have a bottom-left and a top-right endpoint. As a result, a non-blocking edge~$e=\lbrace u,v\rbrace $ can not pass through any gadget square~$r_w$, except the squares~$r_u$ or~$r_v$ since~$e$ would have to cross the blocking edge of~$r_w$.
Accordingly, it is straight-forward to obtain a geodesic drawing of~$\Gamma_g'$: We remove the blocking edges, reinsert the vertices of~$V$ according to the mappings~$x'$ and~$\gamma$ and connect them to the vertices of their respective gadgets in a geodesic fashion. This can always be done without crossings. Figure~\ref{fig:olp-to-gp01}(f) shows one possibility. 
If the edge from~$v_2$ passes to the left of~$v_1$, we may have to choose a reflected version.
Finally, we remove the vertices~$v_1$ and~$v_2$ which now act as subdivision vertices.

On the other hand, let~$\Gamma_g'$ be a geodesic planar drawing of~$\mathcal G_g'$. Without loss of generality, we can assume that each edge~$\lbrace u,v\rbrace$ passes only through the squares of~$u$ and~$v$. Furthermore, for each~$v\in V$ we can assume that its incident edges intersect the boundary of~$r_v$ only to the top-right of $p_v^{tr}-(1/48,1/48)$ or to the bottom-left of $p_v^{bl}+(1/48,1/48)$, see Figure~\ref{fig:olp-to-gp01}(g). 
Thus, we can simply remove the parts of the edges in the interior of the gadget squares and connect the gadget vertices to the intersection points of the edges with the gadget squares in a geodesic fashion.
\end{proofsketch}

The bit size of the numbers involved in the calculations of our reduction is linearly bounded  in the bit size of the directions of~$S$. Together with Theorem~\ref{theorem:2olp-npc} we obtain the proof of Theorem~\ref{theorem:lgp}.
The instances generated by Lemma~\ref{lemma:olp-to-lgp} are in general position. In particular, this means that the mappings~$x$ and~$y$ are injective. We obtain an immediate reduction to \textsc{Bi-Monotonicity}. The correctness follows from the fact that every $L_1$-geodesic rectilinear path can be transformed into a bi-monotone curve and vice versa. Thus, we obtain Theorem~\ref{theorem:BiMon}.

\section{Ordered Level Planarity}
\label{sec:olp}

\begin{figure}[tb]
  \centering
  \subcaptionbox{\label{fig:PM3SAT}}{
    \centering
    \includegraphics[width=0.25\columnwidth, page=1]{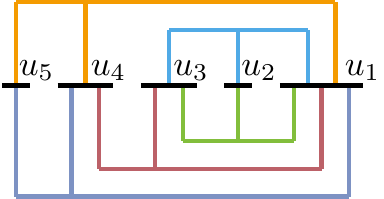}}
  ~~~~~~~~~~
    \subcaptionbox{\label{fig:PM3SAT-2}}{
    \centering
    \includegraphics[width=0.5\columnwidth, page=3]{fig/PM3SAT}}\\
  \subcaptionbox{\label{fig:T_0}}{
    \centering
    \includegraphics[width=0.99\columnwidth, page=5]{fig/PM3SAT}}
  \caption{\textbf{(a)} Representation~$\mathcal R$ of~$\varphi$ with negative clauses $(\overline u_1\vee \overline u_4 \vee \overline u_5 )$, $(\overline u_1 \vee \overline u_3 \vee \overline u_4 )$ and $(\overline u_1 \vee \overline u_2 \vee \overline u_3 )$ and positive clauses $(u_1 \vee u_4 \vee u_5 )$ and $(u_1 \vee u_2 \vee u_3 )$ and \textbf{(b)} its modified version $\mathcal R'$ in Lemma~\ref{theorem:olp-nph}. \textbf{(c)} Tier $\mathcal T_0$.}
\end{figure}

To show $\mathcal{NP}$-hardness of \textsc{Ordered Level Planarity} we reduce from a \textsc{3-Satisfiability} variant described in this paragraph.
A \emph{monotone} \textsc{3-Satisfiability} formula is a Boolean \textsc{3-Satisfiability} formula in which each clause is either \emph{positive} or \emph{negative}, that is, each clause contains either exclusively positive or exclusively negative literals respectively.
A \emph{planar} 3SAT formula $\varphi=(\mathcal U,\mathcal C)$ is a  Boolean \textsc{3-Satisfiability} formula with a set $\U$ of variables and a set $\C$ of clauses such that its \emph{variable-clause graph} $G_\varphi = (\U \uplus \C, E)$ is planar.
The graph~$G_\varphi$ is bipartite, i.e.~every edge in~$E$ is incident to both a \emph{clause} vertex from~$\mathcal C$ and a \emph{variable} vertex from~$\mathcal U$.
Furthermore, edge~$\lbrace c,u\rbrace \in E$ if and only if a literal of variable $u\in \mathcal U$ occurs in $c\in \mathcal C$.
\textsc{Planar Monotone 3-Satisfiability} is a special case of \textsc{3-Satisfiability} where we are given a planar and monotone \textsc{3-Satisfiability} formula~$\varphi$ and a \emph{monotone rectilinear representation}~$\mathcal R$ of the variable-clause graph of~$\varphi$. The representation~$\mathcal R$ is a contact representation on an integer grid in which the variables are represented by horizontal line segments arranged on a line~$\ell$. The clauses are represented by E-shapes turned by $90^\circ$ such that all positive clauses are placed above~$\ell$ and all negative clauses are placed below~$\ell$, see Figure~\ref{fig:PM3SAT}. \textsc{Planar Monotone 3-Satisfiability} is~$\mathcal{NP}$-complete~\cite{DBLP:journals/ijcga/BergK12}.
We are now equipped to prove the core lemma of this section.

\newcommand{\lemmaOlpNph}{\textsc{Planar Monotone 3-Satisfiability} reduces in polynomial time to \textsc{Ordered Level Planarity}. The resulting instances have maximum degree~$\Delta=2$ and all vertices on levels with width at least 3 have out-degree at most $1$ and in-degree at most $1$.}
\begin{lemma}
\label{theorem:olp-nph}
\lemmaOlpNph
\end{lemma}

\begin{proofsketch}
We perform a polynomial-time reduction from \textsc{Planar Monotone 3-Satisfiability}. Let~$\mathcal \varphi=(\mathcal U,\mathcal C)$ be a planar and monotone \textsc{3-Satisfiability} formula with~$\mathcal C=\lbrace c_1,\dots,c_{|\mathcal C|}\rbrace$. Let~$G_\varphi$ the variable-clause graph of~$\varphi$. Let~$\mathcal R$ be a monotone rectilinear representation of~$G_\varphi$.  We construct an ordered level graph $\mathcal G=(G,\gamma,\chi)$ such that~$\mathcal G$ has an ordered level planar drawing if and only if~$\varphi$ is satisfiable.
In this proof sketch we omit some technical details such as precise level assignments and level orderings.

\textbf{Overview:}
The ordered level graph~$\mathcal G$ has~$l_3+1$ levels which are partitioned into four \emph{tiers} $\mathcal T_0=\lbrace 0,\dots ,l_0\rbrace$, $\mathcal T_1=\lbrace l_0+1,\dots ,l_1\rbrace$, $\mathcal T_2=\lbrace l_1+1,\dots ,l_2\rbrace$ and $\mathcal T_3=\lbrace l_2+1,\dots ,l_3\rbrace$.
Each clause~$c_i\in \mathcal C$ is associated with a \emph{clause} edge~$e_i=(c_i^s,c_i^t)$ starting with~$c_i^s$ in tier~$\mathcal T_0$ and ending with~$c_i^t$ in tier~$\mathcal T_2$. The clause edges have to be drawn in a system of tunnels that encodes the \textsc{3-Satisfiability} formula~$\varphi$.
In~$\mathcal T_0$ the layout of the tunnels corresponds directly to the rectilinear representation~$\mathcal R$, see Figure~\ref{fig:T_0}.
For each E-shape there are three tunnels corresponding to the three literals of the associated clause. 
The bottom vertex~$c_i^s$ of each clause edge~$e_i$ is placed such that~$e_i$ has to be drawn inside one of the three tunnels of the E-shape corresponding to~$c_i$.
This corresponds to the fact that in a satisfying truth assignment every clause has at least one satisfied literal.
In tier~$\mathcal T_1$ we merge all the tunnels corresponding to the same literal. We create variable gadgets that ensure that for each variable~$u$ edges of clauses containing~$u$ can be drawn in the tunnel associated with either the negative or the positive literal of~$u$ but not both. This corresponds to the fact that every variable is set to either true or false. Tiers~$\mathcal T_2$ and~$\mathcal T_3$ have a technical purpose.

We proceed by describing the different tiers in detail.
Recall that in terms of realizability, \textsc{Ordered Level Planarity} is equivalent to the generalized version where~$\gamma$ and~$\chi$ map to the reals. For the sake of convenience we will begin by designing~$\mathcal G$ in this generalized setting. It is easy to transform~$\mathcal G$ such that it satisfies the standard definition in a polynomial-time post processing step.

\textbf{Tier 0 and 2, clause gadgets:}
The clause edges $e_i=(c_i^s,c_i^t)$ end in tier~$\mathcal T_2$. It is composed of~$l_2-l_1=|\mathcal C|$ levels each of which contains precisely one vertex.
We assign $\gamma (c_i^t)=l_1+i$.
Observe that this imposes no constraint on the order in which the edges enter~$\mathcal T_2$.

Tier~$\mathcal T_0$ consists of a system of tunnels that resembles the monotone rectilinear representation~$\mathcal R$ of~$G_\varphi= (\U \uplus \C, E)$, see Figure~\ref{fig:T_0}.
Intuitively it is constructed as follows: We take the top part of~$\mathcal R$, rotate it by~$180^\circ$ and place it to the left of the bottom part such that the variables' line segments align, see Figure~\ref{fig:PM3SAT-2}.
We call the resulting representation~$\mathcal R'$.
For each E-shape in~$\mathcal R'$ we create a \emph{clause gadget}, which is a subgraph composed of $11$ vertices that are placed on a grid close to the E-shape, see Figure~\ref{fig:T_0-E}.
The red vertex at the bottom is the lower vertex~$c_i^s$ of the clause edge~$e_i$ of the clause~$c_i$ corresponding to the E-shape.
Without loss of generality we assume the grid to be fine enough such that the resulting ordered level graph can be drawn as in Figure~\ref{fig:T_0} without crossings. Further, we assume that the~$y$-coordinates of every pair of horizontal segments belonging to distinct E-shapes differ by at least~$3$. This ensures that all vertices on levels with width at least 3 have out-degree at most $1$ and in-degree at most $1$ as stated in the lemma.

The clause gadget (without the clause edge) has a \emph{unique} ordered level planar drawing in the sense that for every level~$V_i$ the left-to-right sequence of vertices and edges intersected by the horizontal line~$L_i$ through~$V_i$ is identical in every ordered level planar drawing. This is due to the fact that the order of the top-most vertices $v_1'$, $v_6$, $v_2'$, $v_7$, $v_3'$ and $v_8$ is fixed. We call the line segments $v_1'v_6$, $v_2'v_7$ and $v_3'v_8$ the \emph{gates} of~$c_i$. Note that the clause edge~$e_i$ has to intersect one of the gates of~$c_i$. This corresponds to the fact the at least one literal of every clause has to be satisfied.

\begin{figure}[tb]
  \centering
  \subcaptionbox{\label{fig:T_0-E-rep}}{
    \centering
    \includegraphics[width=0.35\columnwidth,page=3]{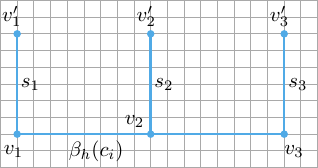}}
  ~~~~~
  \subcaptionbox{\label{fig:T_0-E-graph}}{
    \centering
    \includegraphics[width=0.35\columnwidth,page=4]{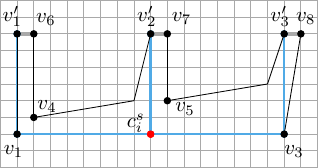}}
  \caption{\textbf{(a)} The E-shape and \textbf{(b)} the clause gadget of clause~$c_i$. The thick gray lines represent the gates of~$c_i$. }
  \label{fig:T_0-E}
\end{figure}

The subgraph~$G_0$ induced by~$\mathcal T_0$ (without the clause edges) has a unique ordered level planar drawing. 
In tier~$\mathcal T_1$ we bundle all gates that belong to one literal together by creating two long paths for each literal. These two paths form the \emph{tunnel} of the corresponding literal. All clause edges intersecting a gate of some literal have to be drawn inside the literal's tunnel, see Figure~\ref{fig:T_0}.
To this end, for~$j=1,\dots,|\mathcal U|$ we use~$N_j^0$ ($n_j^0$) to refer to the left-most (right-most) vertex of a negative clause gadget placed on a line segment of~$\mathcal R'$  representing~$u_j\in \mathcal U$. The vertices~$N_j^0$ and~$n_j^0$ are the first vertices of the paths forming the \emph{negative} tunnel~$T_j^n$ of the negative literal of variable~$u_j$. 
Analogously, we use~$P_j^0$ ($p_j^0$) to refer to the left-most (right-most) vertex of a positive clause gadget placed on a line segment of~$\mathcal R'$ representing~$u_j$.
The vertices~$P_j^0$ and~$p_j^0$ are the first vertices of the paths forming the \emph{positive} tunnel~$T_j^p$ of the positive literal of variable~$u_j$.
If for some~$j$ the variable~$u_j$ is not contained both in negative and positive clauses, we artificially add two vertices ~$N_j^0$ and $n_j^0$ or~$P_j^0$ and~$p_j^0$ on the corresponding line segments  in order to avoid having to treat special cases in the remainder of the construction.

\textbf{Tier 1 and 3, variable gadgets:}
Recall that every clause edge has to pass through a gate that is associated with some literal of the clause, and, thus, every edge is drawn in the tunnel of some literal. We need to ensure that it is not possible to use tunnels associated with the positive, as well as the negative literal of some variable simultaneously. To this end, we create a \emph{variable gadget} with vertices in tier~$\mathcal T_1$ and tier~$T_3$ for each variable. The variable gadget of variable~$u_j$ is illustrated in Figure~\ref{fig:variable_noTunnels}.
The variable gadgets are nested in the sense that they start in~$\mathcal T_1$ in the order~$u_1,u_2,...,u_{|\mathcal U |}$, from bottom to top and they end in the reverse order in~$\mathcal T_3$, see Figure~\ref{fig:nesting}.
We force all tunnels with index at least~$j$ to be drawn between the vertices~$u_j^a$ and~$u_j^b$.
This is done by subdividing the tunnel edges on this level, see Figure~\ref{fig:positive}.
The \emph{long edge}~$(u_j^s,u_j^t)$ has to be drawn to the left or right of~$u_j^c$ in~$\mathcal T_3$.
Accordingly, it is drawn to the left of~$u_j^a$ or to the right of~$u_j^b$ in~$\mathcal T_1$.
Thus, it is drawn either to the right (Figure~\ref{fig:positive}) of all the tunnels or to the left (Figure~\ref{fig:negative}) of all the tunnels.
As a consequence, the \emph{blocking edge}~$(u_j^s,u_j^p)$ is also drawn either to the right or the left of all the tunnels.
Together with the edge~$(u_j^q,u_j^p)$ it prevents clause edges from being drawn either in the positive tunnel~$T_j^p$ or negative tunnel~$T_j^n$ of variable~$u_j$ which end at level~$\gamma(u_j^q)$ because they can not reach their endpoints in~$\mathcal T_2$ without crossings.
We say~$T_j^p$ or~$T_j^n$ are \emph{blocked} respectively.
\begin{figure}[tb]
  \centering
  \subcaptionbox{\label{fig:variable_noTunnels}}{
    \centering
    \includegraphics[width=0.15\columnwidth,page=1]{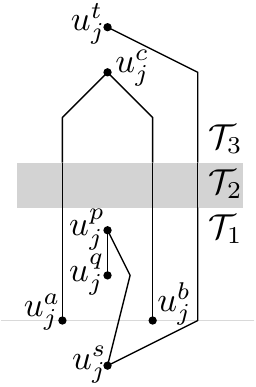}}
  ~~
  \subcaptionbox{\label{fig:positive}}{
    \centering
    \includegraphics[width=0.38\columnwidth,page=2]{fig/variable}}
      ~~
  \subcaptionbox{\label{fig:negative}}{
    \centering
    \includegraphics[width=0.38\columnwidth,page=3]{fig/variable}}
  \caption{\textbf{(a)} The variable gadget of~$u_j$ in \textbf{(b)} positive and \textbf{(c)} negative state.}
\end{figure}

The construction of the ordered level graph~$\mathcal G$ can be carried out in polynomial time. Note that maximum degree is~$\Delta=2$ and that all vertices on levels with width at least 3 have out-degree at most 1 and in-degree at most 1 as claimed in the lemma.

\textbf{Correctness:}
It remains to show that~$\mathcal G$ has an ordered level planar drawing if and only if~$\varphi$ is satisfiable. Assume that~$\mathcal G$ has an ordered level planar drawing~$\Gamma$. We create a satisfying truth assignment for~$\varphi$.
If~$T_j^n$ is blocked we set~$u_j$ to true, otherwise we set~$u_j$ to false for~$j\in 1,\dots ,|\mathcal U|$.
Recall that the subgraph~$G_0$ induced by the vertices in tier~$\mathcal T_0$ has a unique ordered level planar drawing. Consider a clause~$c_i$ and let~$f,g,j$ be the indices of the variables whose literals are contained in~$c_i$. Clause edge~$e_i=(e_i^s,e_i^t)$ has to pass level~$l_0$ through one of the gates of~$c_i$. More precisely, it has to be drawn between either $N_f^0$ and $n_f^0$, $N_g^0$ and $n_g^0$ or $N_j^0$ and $n_j^0$ if~$c_i$ is negative or between either $P_f^0$ and $p_f^0$, $P_g^0$ and $p_g^0$ or $P_j^0$ and $p_j^0$ if~$c_i$ is positive, see Figure~\ref{fig:T_0}. First, assume that~$c_i$ is negative and assume without loss of generality that it traverses~$l_0$ between~$N_j^0$ and~$n_j^0$. In this case clause edge~$e_i$ has to be drawn in $T_j^n$. Recall that this is only possible if~$T_j^n$ is not blocked, which is the case if~$u_j$ is false, see Figure~\ref{fig:negative}. Analogously, if~$c_i$ is positive and~$e_i$ traverses w.l.o.g.~between~$p_j^P$ and~$p_j^p$, then~$u_j$ is true, Figure~\ref{fig:positive}. Thus, we have established that one literal of each clause in~$\mathcal C$ evaluates to true for our truth assignment and, hence, formula~$\varphi$ is satisfiable.
\begin{figure}[tb]
  \centering
    \includegraphics[width=0.99\columnwidth,page=1]{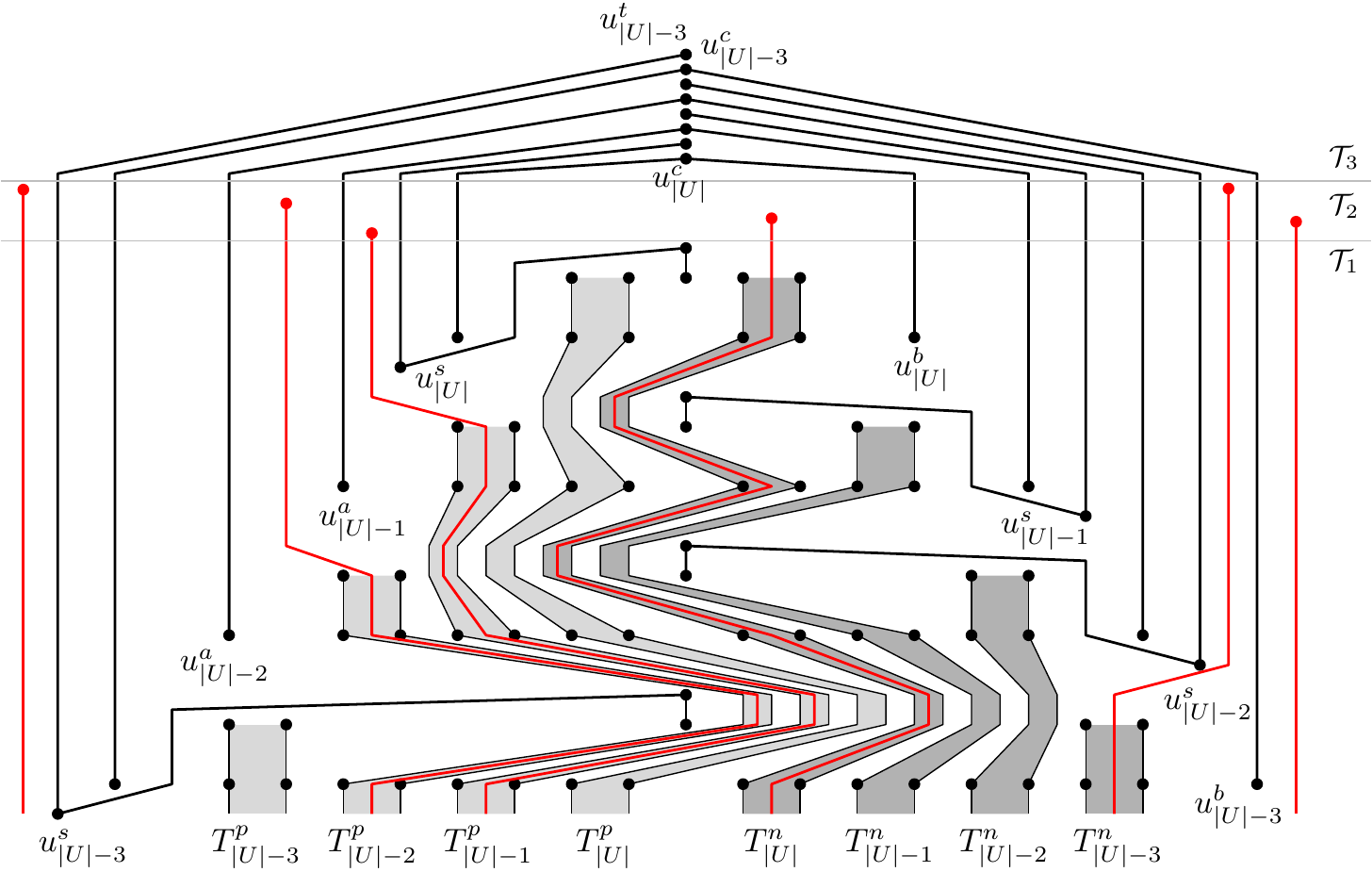}
  \caption{Nesting structure of the variable gadgets.}
  \label{fig:nesting}
\end{figure}

Now assume that~$\varphi$ is satisfiable and consider a satisfying truth assignment. We create an ordered level planar drawing~$\Gamma$ of~$\mathcal G$.
It is clear how to create the unique subdrawing of~$G_0$. The variable gadgets are drawn in a nested fashion, see Figure~\ref{fig:nesting}.
For~$j=1,\dots ,|\mathcal U|-1$ we draw edge~$(u_j^a,u_j^c)$ to left of~$u_{j+1}^a$ and~$u_{j+1}^s$ and edge~$(u_j^b,u_j^c)$ to right of~$u_{j+1}^b$ and~$u_{j+1}^s$. In other words, the pair~$((u_j^a,u_j^c),(u_j^b,u_j^c))$ is drawn between all such pairs with index smaller than~$j$.
Recall that the vertices $u_j^a$, $u_j^b$, $u_j^s$, $u_j^p$ and $u_j^q$ are located on higher levels than the according vertices of variables with index smaller than~$j$ and that $u_j^t$ and $u_j^c$ are located on lower levels than the according vertices of variables with index smaller than~$j$.

For~$j=1,\dots ,|\mathcal U|$ if~$u_j$ is positive we draw the long edge~$(u_j^s,u_j^t)$ to the right of~$u_j^b$ and~$u_j^c$ and, accordingly, we have to draw all tunnels left of~$u_j^s$ and~$u_j^q$ (except for~$T_j^n$, which has to be drawn to the left of~$u_j^s$ and end to the right of~$u_j^q$), see Figure~\ref{fig:positive}.
If~$u_j$ is negative we draw the long edge~$(u_j^s,u_j^t)$ to the left of~$u_j^b$ and~$u_j^c$ and, accordingly, we have to draw all tunnels right of~$u_j^s$ and~$u_j^q$ (except for~$T_j^p$, which has to be drawn to the right of~$u_j^s$ and end to the left of~$u_j^q$), see Figure~\ref{fig:negative}. We have to draw the blocking edge~$(u_j^s,u_j^p)$  to the right of~$n_j^{j+1}$ if~$u_j$ is positive and to the left of~$P_j^{j+1}$ if~$u_j$ is negative.

It remains to describe how to draw the clause edges.
Let $c_i$ be a clause. There is at least one true literal in~$c_i$. Let~$k$ be the index of the corresponding variable. We describe the drawing of clause edge~$e_i=(c_i^s,c_i^t)$ from bottom to top. We start by drawing~$e_i$ in the tunnel~$T_k^p$ ($T_k^n$) if~$c_i$ is positive (negative). After the variable gadget of~$u_k$ the edge~$e_i$ leaves its tunnel and is drawn to the left (right) of all gadgets of variables with higher index, see Figure~\ref{fig:nesting}.
\end{proofsketch}\\

We obtain~$\mathcal{NP}$-hardness for instances with maximum degree~$\Delta=2$. In fact, 
we can  restrict our attention to instances level-width~$\lambda=2$. To this end, we split levels with width~$\lambda_i>2$ into~$\lambda_i-1$ levels containing exactly two vertices each
.

\newcommand{\lemmaOlpToTwo}{An instance~$\mathcal G=(G=(V,E),\gamma,\chi)$ of \textsc{Ordered Level Planarity} with maximum degree~$\Delta \le 2$ can be transformed in linear time into an equivalent instance~$\mathcal G'=(G'=(V',E'),\gamma',\chi')$ of \textsc{Ordered Level Planarity} with level-width~$\lambda' \le 2$ and maximum degree~$\Delta'$. If in~$\mathcal G$ all vertices on levels with width at least 3 have out-degree at most 1 and in-degree at most 1, then~$\Delta'\le 2$. Otherwise,~$\Delta'\le \Delta+1$.}
\begin{lemma}
\label{theorem:olp-to-2olp}
\lemmaOlpToTwo
\end{lemma}

The reduction in Lemma~\ref{theorem:olp-nph} requires degree-2 vertices.
With $\Delta=1$, the problem becomes polynomial-time solvable.
In fact, even if $\Delta=2$ one can easily solve it  as long as the
maximum in-degree and the maximum out-degree are both bounded by 1.
Such instances consists of a set $P$ of $y$-monotone paths.
We write $p \prec q$, meaning that $p\in P$ must be drawn to the left of $q\in P$, if $p$ and $q$ have vertices
$v_p$ and $v_q$ that lie adjacent on a common level.
If $\prec$ is acyclic, we can draw~$\mathcal G$ according to a linear extension of~$\prec$, otherwise there exists no solution
.

\newcommand{\lemmaDegreeOne}{\textsc{Ordered Level Planarity} restricted to instances with maximum in-degree~$\Delta^-=1$ and maximum out-degree~$\Delta^+=1$ can be solved in linear time.}
\begin{lemma}
\label{lemma:degreeOne}
\lemmaDegreeOne
\end{lemma}

For~$\lambda=1$ \textsc{Ordered Level Planarity} is solvable in linear time since \textsc{Level Planarity} can be solved in linear time~\cite{DBLP:conf/gd/JungerLM98}. Proper instances can be solved in linear-time via a sweep through every level. The problem is obviously contained in~$\mathcal{NP}$. The results of this section establish Theorem~\ref{theorem:2olp-npc}.\\

\textbf{Acknowledgements:}
We thank the authors of~\cite{DBLP:conf/gd/KatzKRW09} for providing us with unpublished information regarding their plane sweep approach for \textsc{Manhattan Geodesic Planarity}.\newpage

{\small 
   \bibliographystyle{titto-lncs-01}
   \bibliography{abbrv,bib}
}
\newpage

\appendix

\section{Omitted Proofs in Section~\ref{sec:lgp}}
\label{app:lgp}

\rephrase{Lemma}{\ref{lemma:olp-to-lgp}}{\lemmaOlpToLgp}

\begin{figure}[tb]
  \centering
    \includegraphics[width=0.5\columnwidth,page=2]{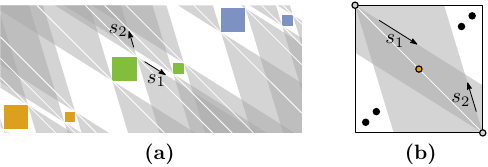}
  \caption{Adaptations (1) \textbf{(a)} and (2) \textbf{(b)} for the general case.}
  \label{fig:olp-to-gp02}
\end{figure}

\begin{proof}
We first prove our claim for the classical case that $S$ contains exactly the four horizontal and vertical directions.
Afterwards, we discuss the necessary adaptations for the general case.
Our reduction is carried out in two steps.
Let $\mathcal G_o=(G_o=(V,E),\gamma,\chi)$ be an \textsc{Ordered Level Planarity} instance with maximum degree~$\Delta =2$ and level-width~$\lambda =2$. 
In Step (i) we turn $\mathcal G_o$ into an equivalent \textsc{Geodesic Planarity} instance $\mathcal G_g'=(G_o,x',\gamma,S)$.
In Step (ii) we transform $\mathcal G_g'$ into an equivalent \textsc{Geodesic Planarity} instance $\mathcal G_g=(G_g,x,y,S)$ where~$G_g$ is a perfect matching and the vertex positions assigned via $x$ and $y$ are in general position.

\textbf{Step (i):}
In order to transform $\mathcal G_o$ into $\mathcal G_g'$ we apply a shearing transformation. We translate the vertices of each level $V_i$ by $3i$ units to the right, see Figure~\ref{fig:olp-to-gp01}(a) and Figure~\ref{fig:olp-to-gp01}(b). Clearly, every geodesic drawing of~$\mathcal G_g'$ can be turned into an ordered level planar drawing of~$\mathcal G_o$. On the other hand, consider an ordered level planar drawing~$\Gamma_o$ of~$\mathcal G_o$. Without loss of generality we can assume that in~$\Gamma_o$ all edges are realized as polygonal paths in which bend points occur only on the horizontal lines~$L_i$  through the levels~$V_i$ where $0\le i\le h$. Further, we may assume that all bend points have $x$-coordinates in the open interval~$(-1,2)$. We shear~$\Gamma_o$ by translating the bend points and vertices of level $V_i$ by $3i$ units to the right for $0\le i\le h$, see Figure~\ref{fig:olp-to-gp01}(b). In the resulting drawing~$\Gamma_o'$, the vertex positions match those of~$\mathcal G_g'$. Furthermore, all edge-segments have a positive slope.
 Thus, since the maximum degree is~$\Delta =2$ we can replace all edge-segments with $L_1$-geodesic rectilinear paths that closely trace the segments and we obtain a geodesic drawing~$\Gamma_g'$ of~$\mathcal G_g'$, see Figure~\ref{fig:olp-to-gp01}(c).

\textbf{The general case:}
It remains to discuss the adaptations for the case that~$S$ is an arbitrary set of directions symmetric with respect to the origin.
By applying a linear transformation we can assume without loss of generality that $(1,0)$ and $(0,1)$ are adjacent directions in~$S$.
Accordingly, all the remaining directions point into the top-left or the bottom-right quadrant.
Further, by vertical scaling we can assume that no direction projects to $(1,1)$ on the unit square.
Observe that if we do not insist on a coordinate assignment in general position, the reduction for the restricted case discussed above is already sufficient. 
In order to guarantee general position we have to avoid points that lie on a line with a direction from~$S$. This requires some easy but a bit technical modifications of our construction.

Note that since no direction of~$S$ points to the top-right or bottom-left quadrant, every pair of \emph{conflicting} vertices from~$\mathcal G_g$ that defines a line parallel to one of the directions in~$S$ has to belong to one or both of the gadgets of two vertices~$u,v\in V$ with~$\gamma(u)=\gamma(v)$.
Let~$s_1$ and~$s_2$ be the flattest and steepest slope of $S\setminus \lbrace (1,0),(0,1)\rbrace$ respectively.
In order to guarantee general position we apply the following two changes.

(1) We increase the horizontal distance in the mapping~$x'$ between each pair of vertices~$u\in V$ and~$v\in V$ with~$\gamma (u)=\gamma (v)$ and~$\chi(u)=0$ and~$\chi(v)=1$ in order to ensure that there can not be any conflicting vertices~$u'$,$v'$ in~$\mathcal G_g$ such that~$u'$ belongs to the gadget of~$u$ and~$v$ belongs to the gadget of~$v$ . It suffices to translate~$v$ and its square to the right such that~$p_v^{bl}$ is to the right of the line with direction~$s_1$ through~$p_u^{tr}$, see Figure~\ref{fig:olp-to-gp02}(a).

(2) In order to ensure that there are no conflicting vertices~$w$, $w'$ that belong to the same gadget square~$r_v$, we change the offset to the gadget square corners from~$\pm (1/48)$ and~$\pm(1/24)$ to~$\pm (c/48)$ and~$\pm(c/24)$ where $0<c<1$ is chosen small enough such that the gadget vertices are placed above the line with direction~$s_1$ through~$r_v^{tl}$, below the line with direction~$s_1$ through~$r_v^{br}$, below the line with direction~$s_2$ through~$r_v^{tl}$ and above the line with direction~$s_2$ through~$r_v^{br}$, see Figure~\ref{fig:olp-to-gp02}(b).\hfill \qed
\end{proof}

\section{Omitted Proofs in Section~\ref{sec:olp}}
\label{app:olp}

\rephrase{Lemma}{\ref{theorem:olp-nph}}{\lemmaOlpNph}

\begin{proof}
We perform a polynomial-time reduction from \textsc{Planar Monotone 3-Satisfiability}. Let~$\mathcal \varphi=(\mathcal U,\mathcal C)$ be a planar and monotone \textsc{3-Satisfiability} formula with~$\mathcal C=\lbrace c_1,\dots,c_{|\mathcal C|}\rbrace$. Let~$G_\varphi$ the variable-clause graph of~$\varphi$. Let~$\mathcal R$ be a monotone rectilinear representation of~$G_\varphi$.  We construct an ordered level graph $\mathcal G=(G,\gamma,\chi)$ such that~$\mathcal G$ has an ordered level planar drawing if and only if~$\varphi$ is satisfiable.

\textbf{Overview:}
The ordered level graph~$\mathcal G$ has~$l_3+1$ levels which are partitioned into four \emph{tiers} $\mathcal T_0=\lbrace 0,\dots ,l_0\rbrace$, $\mathcal T_1=\lbrace l_0+1,\dots ,l_1\rbrace$, $\mathcal T_2=\lbrace l_1+1,\dots ,l_2\rbrace$ and $\mathcal T_3=\lbrace l_2+1,\dots ,l_3\rbrace$.
Each clause~$c_i\in \mathcal C$ is associated with a \emph{clause} edge~$e_i=(c_i^s,c_i^t)$ starting with~$c_i^s$ in tier~$\mathcal T_0$ and ending with~$c_i^t$ in tier~$\mathcal T_2$. The clause edges have to be drawn in a system of tunnels that encodes the \textsc{3-Satisfiability} formula~$\varphi$.
In~$\mathcal T_0$ the layout of the tunnels corresponds directly to the rectilinear representation~$\mathcal R$, see Figure~\ref{fig:T_0}.
For each E-shape there are three tunnels corresponding to the three literals of the associated clause. 
The bottom vertex~$c_i^s$ of each clause edge~$e_i$ is placed such that~$e_i$ has to be drawn inside one of the three tunnels of the E-shape corresponding to~$c_i$.
This corresponds to the fact that in a satisfying truth assignment every clause has at least one satisfied literal.
In tier~$\mathcal T_1$ we merge all the tunnels corresponding to the same literal. We create variable gadgets that ensure that for each variable~$u$ edges of clauses containing~$u$ can be drawn in the tunnel associated with either the negative or the positive literal of~$u$ but not both. This corresponds to the fact that every variable is set to either true or false. Tiers~$\mathcal T_2$ and~$\mathcal T_3$ have a technical purpose.

We proceed by describing the different tiers in detail.
Recall that in terms of realizability, \textsc{Ordered Level Planarity} is equivalent to the generalized version where~$\gamma$ and~$\chi$ map to the reals. For the sake of convenience we will begin by designing~$\mathcal G$ in this generalized setting. It is easy to transform~$\mathcal G$ such that it satisfies the standard definition in a polynomial-time post processing step.

\textbf{Tier 0 and 2, clause gadgets:}
The clause edges $e_i=(c_i^s,c_i^t)$ end in tier~$\mathcal T_2$. It is composed of~$l_2-l_1=|\mathcal C|$ levels each of which contains precisely one vertex.
We assign $\gamma (c_i^t)=l_1+i$.
Observe that this imposes no constraint on the order in which the edges enter~$\mathcal T_2$.

Tier~$\mathcal T_0$ consists of a system of tunnels that resembles the monotone rectilinear representation~$\mathcal R$ of~$G_\varphi= (\U \uplus \C, E)$, see Figure~\ref{fig:T_0}.
Intuitively it is constructed as follows: We take the top part of~$\mathcal R$, rotate it by~$180^\circ$ and place it to the left of the bottom part such that the variables' line segments align, see Figure~\ref{fig:PM3SAT-2}.
We call the resulting representation~$\mathcal R'$.
For each E-shape in~$\mathcal R'$ we create a \emph{clause gadget}, which is a subgraph composed of $11$ vertices that are placed on a grid close to the E-shape, see Figure~\ref{fig:T_0-E}.
The red vertex at the bottom is the lower vertex~$c_i^s$ of the clause edge~$e_i$ of the clause~$c_i$ corresponding to the E-shape.
Without loss of generality we assume the grid to be fine enough such that the resulting ordered level graph can be drawn as in Figure~\ref{fig:T_0} without crossings. Further, we assume that the~$y$-coordinates of every pair of horizontal segments belonging to distinct E-shapes differ by at least~$3$. This ensures that all vertices on levels with width at least 3 have out-degree at most $1$ and in-degree at most $1$ as stated in the lemma.

\textbf{Technical Details}:
In the following two paragraphs, we describe the construction of the clause gadgets in detail.

For every~$i=1,\dots,|\mathcal C|$ where~$c_i$ is negative we create its 11-vertex clause gadget as follows, see Figure~\ref{fig:T_0-E}.
Let~$s_1,s_2,s_3$ be the three vertical line segments of the E-shape representing~$c_i$ in~$\mathcal R'$ where~$s_1$ is left-most and~$s_3$ right-most.
Let~$v_1,v_2,v_3$ be the lower endpoints and~$v_1',v_2',v_3'$ be the upper endpoints of~$s_1,s_2,s_3$, respectively.
We place the tail~$c_i^s$ of the clause edge~$e_i$ of~$c_i$ at~$v_2$.
We create new vertices at $v_1$, $v_3$, $v_1'$, $v_2'$, $v_3'$, $v_4=v_1+(1,1)$, $v_5=v_2+(1,2)$ and at $v_6,v_7,v_8$ which are the lattice points one unit to the right of~$v_1',v_2',v_3'$, respectively.
To simplify notation, we identify these new vertices with their locations on the grid.
We add edges $(v_1, v_1')$, $(v_3, v_8)$, $(v_4, v_6)$, $(v_4, v_2')$, $(v_5, v_7)$ and $(v_5, v_3')$ to~$G$.

As stated above, we can assume without loss of generality that the grid is fine enough such that the resulting ordered level graph can be drawn as in Figure~\ref{fig:T_0} without crossing.
It suffices to assume that the horizontal and vertical distance between any two segment endpoints of~$\mathcal R'$ is at least 3 (unless the endpoints lie on a common horizontal or vertical line).

\textbf{Gates and Tunnels}:
The clause gadget (without the clause edge) has a \emph{unique} ordered level planar drawing in the sense that for every level~$V_i$ the left-to-right sequence of vertices and edges intersected by the horizontal line~$L_i$ through~$V_i$ is identical in every ordered level planar drawing. This is due to the fact that the order of the top-most vertices $v_1'$, $v_6$, $v_2'$, $v_7$, $v_3'$ and $v_8$ is fixed. We call the line segments $v_1'v_6$, $v_2'v_7$ and $v_3'v_8$ the \emph{gates} of~$c_i$. Note that the clause edge~$e_i$ has to intersect one of the gates of~$c_i$. This corresponds to the fact the at least one literal of every clause has to be satisfied.

The subgraph~$G_0$ induced by~$\mathcal T_0$ (without the clause edges) has a unique ordered level planar drawing. 
In tier~$\mathcal T_1$ we bundle all gates that belong to one literal together by creating two long paths for each literal. These two paths form the \emph{tunnel} of the corresponding literal. All clause edges intersecting a gate of some literal have to be drawn inside the literal's tunnel, see Figure~\ref{fig:T_0}.
To this end, for~$j=1,\dots,|\mathcal U|$ we use~$N_j^0$ ($n_j^0$) to refer to the left-most (right-most) vertex of a negative clause gadget placed on a line segment of~$\mathcal R'$  representing~$u_j\in \mathcal U$. The vertices~$N_j^0$ and~$n_j^0$ are the first vertices of the paths forming the \emph{negative} tunnel~$T_j^n$ of the negative literal of variable~$u_j$. 
Analogously, we use~$P_j^0$ ($p_j^0$) to refer to the left-most (right-most) vertex of a positive clause gadget placed on a line segment of~$\mathcal R'$ representing~$u_j$.
The vertices~$P_j^0$ and~$p_j^0$ are the first vertices of the paths forming the \emph{positive} tunnel~$T_j^p$ of the positive literal of variable~$u_j$.
If for some~$j$ the variable~$u_j$ is not contained both in negative and positive clauses, we artificially add two vertices ~$N_j^0$ and $n_j^0$ or~$P_j^0$ and~$p_j^0$ on the corresponding line segments  in order to avoid having to treat special cases in the remainder of the construction.

\textbf{Tier 1 and 3, variable gadgets:}
Recall that every clause edge has to pass through a gate that is associated with some literal of the clause, and, thus, every edge is drawn in the tunnel of some literal. We need to ensure that it is not possible to use tunnels associated with the positive, as well as the negative literal of some variable simultaneously. To this end, we create a \emph{variable gadget} with vertices in tier~$\mathcal T_1$ and tier~$T_3$ for each variable. The variable gadget of variable~$u_j$ is illustrated in Figure~\ref{fig:variable_noTunnels}.
The variable gadgets are nested in the sense that they start in~$\mathcal T_1$ in the order~$u_1,u_2,...,u_{|\mathcal U |}$, from bottom to top and they end in the reverse order in~$\mathcal T_3$, see Figure~\ref{fig:nesting}.
We force all tunnels with index at least~$j$ to be drawn between the vertices~$u_j^a$ and~$u_j^b$.
This is done by subdividing the tunnel edges on this level, see Figure~\ref{fig:positive}.
The \emph{long edge}~$(u_j^s,u_j^t)$ has to be drawn to the left or right of~$u_j^c$ in~$\mathcal T_3$.
Accordingly, it is drawn to the left of~$u_j^a$ or to the right of~$u_j^b$ in~$\mathcal T_1$.
Thus, it is drawn either to the right (Figure~\ref{fig:positive}) of all the tunnels or to the left (Figure~\ref{fig:negative}) of all the tunnels.
As a consequence, the \emph{blocking edge}~$(u_j^s,u_j^p)$ is also drawn either to the right or the left of all the tunnels.
Together with the edge~$(u_j^q,u_j^p)$ it prevents clause edges from being drawn either in the positive tunnel~$T_j^p$ or negative tunnel~$T_j^n$ of variable~$u_j$ which end at level~$\gamma(u_j^q)$ because they can not reach their endpoints in~$\mathcal T_2$ without crossings.
We say~$T_j^p$ or~$T_j^n$ are \emph{blocked} respectively.

\textbf{Technical Details:}
In the following two paragraphs, we describe the construction of the variable gadgets in detail.

Tier~$\mathcal T_3$ has~$l_3-l_2=2\cdot |\mathcal U|$ layers each of which contains precisely one vertex. We refer to the vertex in layer~$(l_3-2j+1)$ as~$u_j^t$ and to the vertex in layer~$(l_3-2j)$ as~$u_j^c$ for~$j=1,\dots ,|\mathcal U|$. Tier~$\mathcal T_1$ has~$l_1-l_0=4\cdot |\mathcal U|$ levels. In each of the levels~$(l_0+4j-3)$,~$(l_0+4j-1)$ and~$(l_0+4j)$ where~$j=1,\dots ,|\mathcal U|$ we create one vertex. These vertices are called~$u_j^s$, $u_j^q$ and~$u_j^p$ respectively.  In level~$(l_0+4j-2)$ we create two vertices~$u_j^a$ and~$u_j^b$ in this order. We add the edges $(u_j^s,u_j^t)$, $(u_j^s,u_j^p)$, $(u_j^a,u_j^c)$, $(u_j^b,u_j^c)$ and~$(u_j^q,u_j^p)$.

Finally, for $j=1,\dots ,|\mathcal U|$ do the following, see Figure~\ref{fig:positive} or Figure~\ref{fig:negative}. In level~$(l_0+4j-2)$ we create vertices $P_j^j, p_j^j,\dots ,P_{|\mathcal U|}^j,p_{|\mathcal U|}^j$, $N_{|\mathcal U|}^j,n_{|\mathcal U|}^j,\dots ,N_j^j, n_j^j$ and add them in this order between $u_j^a$ and $u_j^b$. In level~$(l_0+4j-1)$ we create vertices~$P_j^{j+1}$ and $p_j^{j+1}$ in this order before $u_j^q$ and we create vertices~$N_j^{j+1}$ and $n_j^{j+1}$ in this order after~$u_j^q$.
We create edges realizing the paths $t_j^P=(P_j^0,\dots ,P_j^{j+1})$, $t_j^p=(p_j^0,\dots ,p_j^{j+1})$, $t_j^N=(N_j^0,\dots ,N_j^{j+1})$ and $t_j^n=(n_j^0,\dots ,n_j^{j+1})$.
The pair of paths $T_j^p=(t_j^P,t_j^p)$ is the positive tunnel of variable~$u_j$. The pair of paths $T_j^n=(t_j^N,t_j^n)$ is the negative  tunnel of variable~$u_j$. If an edge~$e$ is drawn in the region between the two paths of a tunnel~$T$, we say it is drawn \emph{in} $T$.

\textbf{Runtime and Properties:}
The construction of the ordered level graph~$\mathcal G$ can be carried out in polynomial time. Note that maximum degree is~$\Delta=2$ and that all vertices on levels with width at least 3 have out-degree at most 1 and in-degree at most 1 as claimed in the lemma.

\textbf{Correctness:}
It remains to show that~$\mathcal G$ has an ordered level planar drawing if and only if~$\varphi$ is satisfiable. Assume that~$\mathcal G$ has an ordered level planar drawing~$\Gamma$. We create a satisfying truth assignment for~$\varphi$.
If~$T_j^n$ is blocked we set~$u_j$ to true, otherwise we set~$u_j$ to false for~$j\in 1,\dots ,|\mathcal U|$.
Recall that the subgraph~$G_0$ induced by the vertices in tier~$\mathcal T_0$ has a unique ordered level planar drawing. Consider a clause~$c_i$ and let~$f,g,j$ be the indices of the variables whose literals are contained in~$c_i$. Clause edge~$e_i=(e_i^s,e_i^t)$ has to pass level~$l_0$ through one of the gates of~$c_i$. More precisely, it has to be drawn between either $N_f^0$ and $n_f^0$, $N_g^0$ and $n_g^0$ or $N_j^0$ and $n_j^0$ if~$c_i$ is negative or between either $P_f^0$ and $p_f^0$, $P_g^0$ and $p_g^0$ or $P_j^0$ and $p_j^0$ if~$c_i$ is positive, see Figure~\ref{fig:T_0}. First, assume that~$c_i$ is negative and assume without loss of generality that it traverses~$l_0$ between~$N_j^0$ and~$n_j^0$. In this case clause edge~$e_i$ has to be drawn in $T_j^n$. Recall that this is only possible if~$T_j^n$ is not blocked, which is the case if~$u_j$ is false, see Figure~\ref{fig:negative}. Analogously, if~$c_i$ is positive and~$e_i$ traverses w.l.o.g.~between~$p_j^P$ and~$p_j^p$, then~$u_j$ is true, Figure~\ref{fig:positive}. Thus, we have established that one literal of each clause in~$\mathcal C$ evaluates to true for our truth assignment and, hence, formula~$\varphi$ is satisfiable.

Now assume that~$\varphi$ is satisfiable and consider a satisfying truth assignment. We create an ordered level planar drawing~$\Gamma$ of~$\mathcal G$.
It is clear how to create the unique subdrawing of~$G_0$. The variable gadgets are drawn in a nested fashion, see Figure~\ref{fig:nesting}.
For~$j=1,\dots ,|\mathcal U|-1$ we draw edge~$(u_j^a,u_j^c)$ to left of~$u_{j+1}^a$ and~$u_{j+1}^s$ and edge~$(u_j^b,u_j^c)$ to right of~$u_{j+1}^b$ and~$u_{j+1}^s$. In other words, the pair~$((u_j^a,u_j^c),(u_j^b,u_j^c))$ is drawn between all such pairs with index smaller than~$j$.
Recall that the vertices $u_j^a$, $u_j^b$, $u_j^s$, $u_j^p$ and $u_j^q$ are located on higher levels than the according vertices of variables with index smaller than~$j$ and that $u_j^t$ and $u_j^c$ are located on lower levels than the according vertices of variables with index smaller than~$j$.

For~$j=1,\dots ,|\mathcal U|$ if~$u_j$ is positive we draw the long edge~$(u_j^s,u_j^t)$ to the right of~$u_j^b$ and~$u_j^c$ and, accordingly, we have to draw all tunnels left of~$u_j^s$ and~$u_j^q$ (except for~$T_j^n$, which has to be drawn to the left of~$u_j^s$ and end to the right of~$u_j^q$), see Figure~\ref{fig:positive}.
If~$u_j$ is negative we draw the long edge~$(u_j^s,u_j^t)$ to the left of~$u_j^b$ and~$u_j^c$ and, accordingly, we have to draw all tunnels right of~$u_j^s$ and~$u_j^q$ (except for~$T_j^p$, which has to be drawn to the right of~$u_j^s$ and end to the left of~$u_j^q$), see Figure~\ref{fig:negative}. We have to draw the blocking edge~$(u_j^s,u_j^p)$  to the right of~$n_j^{j+1}$ if~$u_j$ is positive and to the left of~$P_j^{j+1}$ if~$u_j$ is negative.

It remains to describe how to draw the clause edges.
Let $c_i$ be a clause. There is at least one true literal in~$c_i$. Let~$k$ be the index of the corresponding variable. We describe the drawing of clause edge~$e_i=(c_i^s,c_i^t)$ from bottom to top. We start by drawing~$e_i$ in the tunnel~$T_k^p$ ($T_k^n$) if~$c_i$ is positive (negative). Immediately after level~$\gamma (p_k^{k+1})$ we end up to the left (right) of all tunnels with index larger than~$k$, see Figure~\ref{fig:positive} (Figure~\ref{fig:negative}). Note that since~$T_k^p$ ($T_k^n$) is not blocked we can continue without having to cross blocking edge~$(u_k^s,u_k^p)$ or~$(u_k^q,u_k^p)$. We  draw~$e_i$ to the left (right) of all vertices belonging to variable gadgets with index larger than~$k$, see Figure~\ref{fig:nesting}.
This introduces no crossings since above level~$\gamma (p_k^{k+1})$ all tunnels with index larger than~$k$ are drawn to the 
right of ~$u_{k+1}^a,\dots ,u_{|\mathcal U|}^a$ and the left of $u_{k+1}^b,\dots ,u_{|\mathcal U|}^b$. Connecting to~$c_i^t$ in tier~$\mathcal T_2$ is straight-forward since every level contains only one vertex.\hfill \qed
\end{proof}

\rephrase{Lemma}{\ref{theorem:olp-to-2olp}}{\lemmaOlpToTwo}

\begin{proof}
Figure~\ref{fig:olp-to-2olp} illustrates the following process. For each level~$i$ with~$|V_i|>2$ we replace level~$V_i$ by~$|V_i|-1$ levels with~$2$ vertices each. Accordingly we increase the level of all vertices with a level larger than~$i$ by~$|V_i|-2$. Let~$v_1,\dots ,v_{|V_i|}\in V_i$ with~$\chi(v_1)<\dots<\chi(v_{|V_i|})$. We increase the level of vertex~$v_j$ by~$j-2$ for~$j=3,\dots ,|V_i|$. For $j=2,\dots,|V_i|-1$ we create a vertex~$v_j'$ one level above~$v_j$ with $\chi(v_j')=0$ and we create edge~$(v_j,v_j')$. We call these new edges the \emph{stretch} edges of level~$i$. For~$j=2,\dots,|V_i|$ we set~$\chi (v_j)=1$. For~$j=2,\dots,|V_i|-1$ and for every edge~$(v_j,t)\in E$ we replace~$(v_j,t)$ with~$(v_j',t)$. Let~$\mathcal G'$ denote the resulting instance, which can be constructed in linear time. Observe that the vertex degrees behave as desired.

Clearly, if~$\mathcal G$ has an ordered level planar drawing, then~$\mathcal G'$ has an ordered level planar drawing.
On the other hand, if~$\mathcal G'$ has an ordered level planar drawing, then~$\mathcal G$ also has an ordered level planar drawing: The subgraph induced by the stretch edges of any level has a unique ordered level planar drawing. Further, if an edge~$e$ is drawn right of some~$v_j$ and left of~$v_{j+1}$ then it has to be drawn to the right of~$v_j'$ and left of~$v_{j+1}'$ due to the stretch edges. Therefore, we can transform an ordered level planar drawing of~$\mathcal G'$ into a drawing of~$\mathcal G$ essentially by contracting the stretch edge to single vertices and by removing the resulting superfluous levels.\hfill \qed
\end{proof}

\begin{figure}[tb]
  \centering
  \subcaptionbox{\label{fig:olp-to-2olp}}{
    \centering
    \includegraphics[page=1]{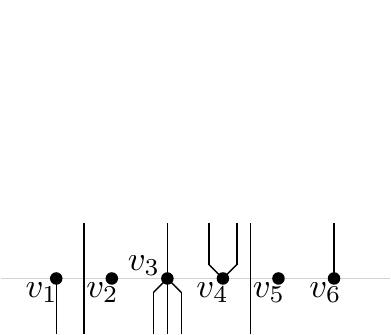}}
  ~~~~~~~~~~
  \subcaptionbox{\label{fig:olp-to-2olp}}{
    \centering
    \includegraphics[page=2]{fig/olp-to-2olp}}
  \caption{In order to reduce from Ordered Level Planarity each level $i$ with $k>2$ vertices (Fig.~\textbf{(a)}) is replaced with $k-1$ levels (Fig.~\textbf{(b)}). Thick edges are the stretch edges of level~$i$.  }
  \label{fig:olp-to-2olp}
\end{figure}

\rephrase{Lemma}{\ref{lemma:degreeOne}}{\lemmaDegreeOne}

\begin{proof}
Let $\mathcal G=(G=(V,E),\gamma,\chi)$ be an ordered level graph with maximum indegree~$\Delta^-=1$ and maximum outdegree~$\Delta^+=1$.
Such a graph~$\mathcal G$ consists of a set $P$ of
$y$-monotone paths. Each path $p\in P$ has vertices on some sequence of
levels, possibly skipping intermediate levels.

We define the following relation on $P$:
We write $p \prec q$, meaning that $p$ must be drawn to the left of $q$, if $p$ and $q$ have vertices
$v_p$ and $v_q$ that lie adjacent on a common level, i.e.~$
\gamma(v_p)=\gamma(v_q)$ and
$\chi(v_q) = \chi(v_p)+1$.
This relation has at most $|V|$ pairs, and by topological sorting,
we can find in $O(|V|)$ time a linear ordering
 that
is consistent with the relation $\prec$, unless this relation has a
cycle.
The former case implies the existence of an ordered level drawing while the  latter case implies that the problem has no solution.

This follows from considerations about horizontal separability of
$y$-monotone sets by translations,
cf.~\cite{deB,amr-qpscu-15}. An easy proof can be given following
Guibas and Yao~\cite{gy-tsr-80,gy-tsr-83}:
Consider an ordered level planar drawing of~$\mathcal G$.
Among the paths whose lower endpoint is visible from the left, the one with the topmost lower endpoint must precede all other paths
to which it is related in the $\prec$-relation. Removing this path
and iterating the procedure leads to a linear order that
extends~$\prec$.
On the other hand, if we have such a linear order
 $x\colon
P\to\{1,\ldots,|P|\}$, we can simply draw each path $p$ straight at
$x$-coordinate $x(p)$, subdivide all edges properly and, finally, shift the vertices on each level such that the vertices of~$V$ are placed according to~$\chi$ while maintaining the order~$x$.
\hfill \qed
\end{proof}

\section{Relationship to Level Planarity Variants}
\label{sec:clusterAndT}

\textbf{Clustered Level Planarity:}
Forster and Bachmaier~\cite{DBLP:conf/sofsem/ForsterB04} introduced a version of \textsc{Level Planarity} that allows the visualization of vertex clusterings. A \emph{clustered} level graph~$\mathcal G$ is a triple~$(G=(V,E),\gamma,T)$ where~$(G,\gamma)$ is a level graph and $T$ is a \emph{cluster hierarchy}, i.e. a rooted tree whose leaves are the vertices in~$V$. Each internal node of~$T$ is called \emph{cluster}. The \emph{vertices} of a cluster~$c$ are the leaves of the subtree of~$T$ rooted at~$c$.
A cluster hierarchy is \emph{flat} if all leaves have distance at most two from the root.
A \emph{clustered} level planar drawing of a clustered level graph~$\mathcal G$ is a level planar drawing of~$(G,\gamma)$ together with a closed simple curve for each cluster that encloses precisely the vertices of the cluster such that the following conditions hold:
(i) no two cluster boundaries intersect (ii) every edge crosses each cluster boundary at most once (iii) the intersection of any cluster with the horizontal line~$L_i$ through level~$V_i$ is either a line segment or empty for any level~$V_i$.
The problem \textsc{Clustered Level Planarity} asks whether a given clustered level graph has a clustered level planar drawing. Forster and Bachmaier~\cite{DBLP:conf/sofsem/ForsterB04} presented a $O(h|V|)$  algorithm for a special case of proper clustered level graphs, where~$h$ is the height of~$\mathcal G$. Angelini, Da Lozzo, Di Battista, Frati, and Roselli~\cite{DBLP:journals/tcs/AngeliniLBFR15} provided a quartic-time algorithm for all proper instances. The general version of \textsc{Clustered Level Planarity} is $\mathcal{NP}$-complete~\cite{DBLP:journals/tcs/AngeliniLBFR15}.

\textbf{T-Level Planarity:}
This variation of \textsc{Level Planarity} considers consecutivity constraints for the vertices on each level. A \emph{$\mathrm{T}$-level} graph~$\mathcal G$ is a triple~$(G=(V,E),\gamma,\mathcal T)$ where~$(G,\gamma)$ is a level graph and $\mathcal T=(T_0,\dots ,T_{h})$ is a set of trees where the leaves of~$T_i$ are $V_i$.
An \emph{$\mathrm T$-level planar} drawing of a $\mathrm T$-level graph~$\mathcal G$ is a level planar drawing of~$(G,\gamma)$ such that for every level $V_i$ and for each node $u$ of~$T_i$ the leaves of the subtree of~$T_i$ rooted at~$u$ appear consecutively along~$L_i$. The problem \textsc{T-Level Planarity} asks whether a given $\mathrm T$-level graph has a $\mathrm T$-level planar drawing. Wotzlaw, Speckenmeyer and Porschen~\cite{DBLP:journals/dam/WotzlawSP12} introduced the problem and provided a quadratic-time algorithm for proper instances with constant level-width. Angelini et al.~\cite{DBLP:journals/tcs/AngeliniLBFR15} give a quartic-time algorithm for proper instances with unbounded level-width. For general $\mathrm T$-level graphs the problem is $\mathcal{NP}$-complete~\cite{DBLP:journals/tcs/AngeliniLBFR15}.\\

\rephrase{Theorem}{\ref{theorem:olp-to-tlp}}{\theoremTlp}

\begin{figure}[tb]
  \centering
  \subcaptionbox{\label{fig:tlp}}{
    \centering
    \includegraphics[width=0.2\columnwidth,page=4]{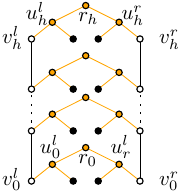}}
  ~~~~~
  \subcaptionbox{\label{fig:clp01}}{
    \centering
    \includegraphics[width=0.062\columnwidth,page=14]{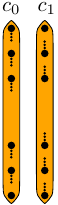}}
      ~~~~~
  \subcaptionbox{\label{fig:clp02}}{
    \centering
    \includegraphics[width=0.25\columnwidth,page=15]{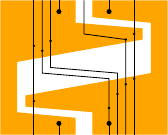}}
  \caption{Reductions from \textsc{Ordered Level Planarity} to \textsc{T-level Planarity} \textbf{(a)} and \textsc{Clustered Level Planarity}~\textbf{(b)}. Big black vertices are the vertices of the \textsc{Ordered Level Planarity} instance. White vertices belong to the paths $p_l$ and $p_r$. In \textbf{(a)} the orange vertices and edges illustrate each level's tree. In \textbf{(b)} and \textbf{(c)} the clusters are represented as orange regions. The small black vertices are subdivision vertices. \textbf{(c)} Subdivided edges can be drawn to the left, to the right or between the two big black vertices while intersecting each cluster boundary at most once.}
\end{figure}

\begin{proof}
Let $\mathcal G=(G=(V,E),\gamma,\pi)$ be an ordered level graph with maximum degree $\Delta =2$ and level-width $\lambda =2$. We augment each level~$V_i$ with~$|V_i|=1$ by adding an isolated dummy vertex~$v$ with $\gamma (v)=i$ and $\chi (v)=1$ in order to avoid having to treat special cases. Thus, each level~$V_i$ has a vertex~$v_i^0$ with~$\chi (v_i^0)=0$ and a vertex~$v_i^1$ with~$\chi(v_i^1)=1$. The following steps are illustrated in Figure~\ref{fig:tlp}. For each level~$V_i$ we create two new vertices~$v_i^l$ and~$v_i^r$. We add edges $(v_i^l,v_{i+1}^l)$ and $(v_i^r,v_{i+1}^r)$ for $i=0,\dots ,h-1$, where~$h$ is the height of~$\mathcal G$. Hence, we obtain a path $p_l$ from $v_0^l$ to $v_h^l$ and a path $p_r$ from $v_0^r$ to $v_h^r$. The root $r_i$ of each tree $T_i$ has two children $u_i^l$ and $u_i^r$. The two children of~$u_i^l$ are~$v_i ^l$ and $v_i^0$. The two children of~$u_i^r$ are~$v_i ^r$ and~$v_i^1$. Let~$\mathcal G'$ denote the resulting $\mathrm{T}$-level graph. The construction of~$\mathcal G'$ can clearly be carried out in linear time.

Clearly, an ordered level planar drawing~$\Gamma$ of~$\mathcal G$ can be augmented to a $\mathrm{T}$-level planar drawing of~$\mathcal G'$ by drawing~$p_l$ to the left of~$\Gamma$ and by drawing~$p_r$ to the right of~$\Gamma$. On the other hand, let~$\Gamma'$ be a $\mathrm{T}$-level-planar drawing of~$\mathcal G'$. We can assume without loss of generality that all vertices are placed on vertical lines with $x$-coordinate $-1$, $0$, $1$ or $2$. The paths~$p_l$ and~$p_r$ are vertex-disjoint and drawn without crossing. Thus, $p_l$ is drawn either to the left or to the right of~$p_r$. 
By reflecting horizontally at the line~$x=1/2$ we can assume without loss of generality that~$p_l$ is drawn to the left of~$p_r$.
Consequently, for each level~$V_i$ the vertex $v_i^0$ has to be drawn to the left of the vertex~$v_i^1$ since~$v_i^l$ and~$v_i^0$ are the children of~$u_i^l$ and since $v_i^r$ and~$v_i^1$ are the children of~$u_i^r$. Therefore, the subdrawing of~$G$ or its mirror image is an ordered level planar drawing of~$\mathcal G$.\hfill \qed
\end{proof}

\rephrase{Theorem}{\ref{theorem:olp-to-clp}}{\theoremClp}

\begin{proof}
Let $\mathcal G=(G=(V,E),\gamma,\pi)$ be an ordered level graph with maximum degree $\Delta =2$ and level-width $\lambda =2$. We augment each level~$V_i$ with~$|V_i|=1$ by adding an isolated dummy vertex~$v$ with $\gamma (v)=i$ and $\chi (v)=1$. Thus, each level~$V_i$ has a vertex~$v_i^0$ with~$\chi (v_i^0)=0$ and a vertex~$v_i^1$ with~$\chi(v_i^1)=1$. The following steps are illustrated in Figure~\ref{fig:clp01}. 
In addition to the trivial cluster that contains all vertices, we create two clusters~$c_0$ and~$c_1$. Cluster~$c_0$ contains the vertices~$v_i^0$ and cluster~$c_1$ contains the vertices~$v_i^1$ for $i=0,\dots, h$.

By subdivision, we transform every edge from some level~$i$ to some level~$j$ into a path of $2(j-i)+1$ edges. This path will alternatively enter~$c_0$ and~$c_1$ but each subdivision edge crosses the boundary of each cluster at most once.
More precisely, for each level index~$i=0,\dots,h-1$ we do the following. We subdivide each edge~$(u,v)\in E$ with~$\gamma(u)\le i$ and~$\gamma (v)\ge i+1$ twice. The upper of the resulting subdivision vertices is added to~$c_0$, the lower to~$c_1$. The subdivision vertices added to~$c_0$ are placed on new distinct levels added between~$V_i$ and~$V_{i+1}$. Below these new levels and above~$V_i$ we place the subdivision vertices added to~$c_1$,  again on new distinct levels, see Figure~\ref{fig:clp01}. Note that the realizability of~$\mathcal G$ as an ordered level planar drawing is invariant under the described subdivision since every subdivision vertex is the singleton vertex of some new level, see Figure~\ref{fig:clp02}. Let~$\mathcal G'$ denote the resulting  clustered level graph. Since edges may stretch over a linear number of levels, the construction of~$\mathcal G'$ can increase the size of the graph quadratically and, therefore, may require quadratic time.

It is straight-forward to augment an ordered level planar drawing~$\Gamma$ of~$\mathcal G^s$ to create a clustered level planar drawing of~$\mathcal G'$, where~$\mathcal G^s=(G^s,\gamma^s,\chi^s)$ is the ordered level graph obtained by applying the described edge subdivision to~$\mathcal G$.
To this end, we simply draw the cluster's curve appropriately. In particular, the subdivision vertices allow us to maintain the property that edges do not traverse a cluster's boundary more than once, see Figure~\ref{fig:clp02}.

For the other direction, let~$\Gamma'$ be a clustered level planar drawing of~$\mathcal G'$. We can assume without loss of generality that all vertices are placed on vertical lines with $x$-coordinate $0$ or $1$. 
The two clusters pass through every level, their boundaries are not allowed to intersect and they can not be nested. Thus, by reflecting horizontally at the line~$x=1/2$ we can assume without loss of generality that~$c_0$ intersects each level to the left of~$c_1$. Consequently, on each level~$V_i$ the vertex~$v_i^0\in c_0$ is placed to the left of~$v_i^1\in c_1$. Therefore, the subdrawing of~$G^s$ or its mirror image is an ordered level planar drawing of~$\mathcal G^s$.\hfill \qed
\end{proof}

\end{document}